\pdfoutput=1
\documentclass[fleqn,usenatbib,usedcolumn]{mnras}
\usepackage[british]{babel}             
\usepackage{mathptmx}

\usepackage[T1]{fontenc}

\usepackage{graphicx}	
\usepackage{amsmath}	
\usepackage{amssymb}	

\newcommand{\ofe}{\ensuremath{[\textrm{O}/\textrm{Fe}]}}
\newcommand{\nafe}{\ensuremath{[\textrm{Na}/\textrm{Fe}]}}
\newcommand{\cfe}{\ensuremath{[\textrm{C}/\textrm{Fe}]}}
\newcommand{\nfe}{\ensuremath{[\textrm{N}/\textrm{Fe}]}}
\newcommand{\feh}{\ensuremath{[\textrm{Fe}/\textrm{H}]}}
\newcommand{\bafe}{\ensuremath{[\textrm{Ba}/\textrm{Fe}]}}
\newcommand{\sife}{\ensuremath{[\textrm{Si}/\textrm{Fe}]}}
\newcommand{\mgfe}{\ensuremath{[\textrm{Mg}/\textrm{Fe}]}}
\newcommand{\sch}{\ensuremath{\textrm{S}_2(\textrm{CH})}}
\newcommand{\hkdash}{\ensuremath{\textrm{HK}^\prime}}
\newcommand{\chold}{\ensuremath{\textrm{CH}(4300)}}
\newcommand{\cnblue}{\ensuremath{\textrm{S}(3839)}}

\newcommand{\dcnblue}{\ensuremath{\delta\textrm{S}(3839)}}
\newcommand{\cno}{\ensuremath{\textrm{C}+\textrm{N}+\textrm{O}}}
\newcommand{\teff}{\ensuremath{\textrm{T}_\textrm{eff}}}
\newcommand{\logg}{\ensuremath{\log \textrm{g}}}

\title[Multiple populations of NGC~1851]{A broad perspective on multiple abundance populations in the globular cluster NGC~1851}

\author[J. D. Simpson et al.]{
Jeffrey D. Simpson,$^{1}$\thanks{E-mail: jeffrey.simpson@aao.gov.au}, Sarah L. Martell$^{2}$ and Colin A. Navin$^{3}$
\\
$^{1}$Australian Astronomical Observatory, North Ryde, NSW 2113, Australia\\
$^{2}$School of Physics, University of New South Wales, Sydney NSW 2052, Australia\\
$^{3}$Department of Physics and Astronomy, Macquarie University, Sydney, NSW 2109, Australia\\
}

\date{Accepted 2016 October 26. Received 2016 October 24; in original form 2016 April 26}

\pubyear{2016}

\begin{document}
\label{firstpage}
\pagerange{\pageref{firstpage}--\pageref{lastpage}}
\maketitle

\begin{abstract}
We present an analysis of the multiple stellar populations of the globular cluster NGC~1851. We used lower resolution spectra of giant stars to  measure CN, CH, and calcium H \& K spectral indices, and determine elemental abundances for carbon and nitrogen. The CN and CH indices were used to confirm that there are four populations of stars in the cluster. The primordial population of stars, with the lowest CN, was found to be generally chemically distinct in elemental abundances from the second generation populations. As expected, \nfe\ increases with increasing CN strength, but the only other element that correlated with CN was barium. The two largest populations of stars were found to have the same rate of carbon astration as the stars ascend the giant branch. We were also able to confirm that four previously identified extratidal stars are chemically associated with the cluster. This work shows the benefit of considering the chemistry of globular clusters with both high- and low-resolution spectra.
\end{abstract}

\begin{keywords}
globular clusters: individual: NGC~1851
\end{keywords}



\section{Introduction}\label{sec:introduction}
The current picture of a typical Galactic globular cluster is that it exhibits star-to-star variations in light elements (e.g., C, N, O, Na), but is homogenous for heavier elements (e.g., Fe, Ca). As with any broad category of astronomical objects, there are exceptions. At one extreme we have objects like Ruprecht 106 \citep{Villanova2013} and E3 \citep{Salinas2015} which could be considered examples of the previous paradigm of globular clusters: simple stellar populations, exhibiting neither bimodality in molecular band strengths nor variation in the light element abundances. At the other extreme we have the well-known examples of objects such as $\omega$~Cen \citep[e.g.,][]{Johnson2010,Marino2012,Simpson2012} and M54 \citep[e.g.,][]{Sarajedini1995,Carretta2010a}, which have abundances ranges in almost every element investigated, including iron-peak and s-process elements, which is not observed in `normal' globular clusters. It is within this latter category that NGC~1851 can be placed.

NGC~1851 has been the target of a large number of photometric and spectroscopic studies of all stellar evolutionary sequences.

\citet{Stetson1981} detected peculiarities in the horizontal branch (HB) with a dense clump of stars at the red end and a hint of another clump at the blue end. This was confirmed by \citet{Walker1992}. \citet{Gratton2012} carried out a spectroscopic study of 91 HB stars and found most red HB (RHB) stars were O-rich and Na-poor with $\sim$10-15\% O-poor and Na-rich; most of the latter group were also Ba-rich. They also derived Ba abundances and some upper limits for N for RHB stars. They determined He and N abundances and detected the characteristic Na-O anticorrelation among blue HB (BHB) stars, with BHB stars generally being more Na-rich and O-poor than the RHB stars. As the BHB stars did not show exceptionally high N they inferred that a large spread in \cno\ was unlikely.

Splits or spreads in the red giant branch (RGB) were found by \citet{Calamida2007} in Str\"omgren photometry, \citet{Han2009} in UVI photometry (U-I colour but not V-I), \citet{Lee2009} in Ca \textit{uvby} photometry and \citet{Cummings2014} in Washington photometry. \citet{Hesser1982} found three CN-rich bright giants ($\sim$18\% of the observed RGB stars) in a spectroscopic study, but their data did not enable them to separate the individual contributions of C and N. \citet{Yong2008a} analyzed the spectra of eight giants and found an Na-O anticorrelation and a large abundance spread of the s-process elements Zr and La that correlated with Al and anticorrelated with O. They also found a small (< 0.1 dex) spread in [Fe/H], but were not satisfied that it was physical given the small sample and the limited set of Fe lines available in their data. A spectroscopic study of four RGB stars by \citet{Yong2009} found \cno\ varied by a factor of four and that Na, Al, Zr and La abundances correlated with \cno. In a large spectroscopic study of 120 RGB stars \citet{Carretta2010} found a metallicity spread consistent with the presence of two groups with a 0.06--0.08 dex difference in metallicity. Both groups showed an Na-O anticorrelation, as well a spread in s-process elemental abundances. A spectroscopic analysis of 15 RGB stars by \citet{Villanova2010} yielded \cno, Na, alpha, iron-peak and s-process element abundances. Red and blue groups separated on a \textit{v} vs \textit{v-y} Str\"omgren CMD did not reveal any differences in either alpha, iron-peak or total \cno\ abundances between the groups. \citet{Campbell2012} found a complex quadrimodal distribution of CN molecular line strengths of RGB and asymptotic giant branch (AGB) stars. \citet{Yong2014b}, in a spectroscopic study of 15 RGB stars, found  a similar result to their earlier study with \cno\ varying by a factor of four with redder RGB stars having a higher \cno\ content. They also found a small iron spread, but it was within their measurement uncertainties.

A split of the sub giant branch (SGB) into fainter (f-SGB) and brighter (b-SGB) populations, in roughly equal numbers, was found in HST ACS data by \citet{Milone2008}. It was also detected in the previously mentioned photometric study of \citet{Han2009}. It was clear in this study that the f-SGB was connected to the redder RGB sequence and hence the f-SGB stars were the progenitors of the redder RGB stars. \citet{Zoccali2009} used VI photometry to determine that the f-SGB was concentrated in the cluster core. The previously mentioned \citet{Cummings2014} study also found a split in the SGB using washington photometry. A split SGB in the cluster core was detected for the first time by \citet{Turri2015}. \citet{Lardo2012} derived C and N (but not O) abundances for 64 MS and SGB stars. They found strongly anticorrelated variations in [C/H] and [N/H] in both the f-SGB and b-SGB stars, with the f-SGB stars showing $\sim$2.5 times the C+N of the b-SGB stars. Using literature Str\"omgren photometric data they also connected the f-SGB with the redder RGB sequence. \citet{Gratton2012a}, in a spectroscopic study of 77 SGB stars, found the f-SGB stars were 0.05 dex more metal-rich and had higher s-process (Sr and Ba) abundances than the b-SGB stars.

The previously mentioned \citet{Campbell2012} study found the same complex distribution of CN line strengths in AGB stars as in the RGB sequence.

Several of the previously mentioned studies also found evidence of multiple populations on the MS, photometrically by \citet{Cummings2014} and in the spectroscopic study of \citet{Lardo2012}.

So what is the overall picture of NGC~1851 that emerges from these studies? It is currently unclear if there is in fact an overall metallicity spread in the cluster. \citet{Carretta2010} detected a 0.06--0.08 dex metallicity spread amongst 120 RGB stars, but \citet{Yong2008a} and \citet{Yong2014b} do not find significant evidence for this.

For the SGB three main scenarios have been advanced to explain the split: (1) the f-SGB is $\sim$1 Gyr older than the b-SGB, given similar overall [Fe/H] metallicity for both populations \citep{Milone2008}; (2) the f-SGB is $\sim$2 Gyr older than the b-SGB and the \cno\ sum of the f-SGB is less than that of the b-SGB by a factor $\sim$2 \citep{Cassisi2008};  (3) the f-SGB is coeval (or slightly younger $\sim$ 300-400 Myr) than the b-SGB with \cno\ enhanced by a factor $\sim$2 compared with the b-SGB \citep[]{Cassisi2008, Ventura2009}.

There is now evidence for \cno\ enhancement of the f-SGB \citep{Lardo2012} and the redder RGB populations \citep[][but see \citealt{Villanova2010,Gratton2012}]{Yong2009, Yong2014b}, so it seems that the latter is the most likely scenario. The f-SGB stars are the progenitors of the redder RGB stars, and this is also consistent with their connection on CMDs \citep[]{Han2009, Lardo2012}. The finding by \citet{Zoccali2009} that the f-SGB population is concentrated in the cluster core, where the enriched gas from a first generation would accumulate, also supports this scenario.

There are two CN anticorrelation sequences with the f-SGB/redder RGB stars enhanced in \cno\ by a factor $\sim$2-3 compared to the b-SGB/bluer RGB stars. The \cno\ abundance appears correlated with the abundance of light elements and s-process elements \citep[]{Yong2008a, Yong2009}. \cno\ is expected to remain constant if fast-rotating massive stars are responsible for the light element and \cno\ abundance variations, so this supports the idea that AGBs are the site of the hot hydrogen burning process that result in the variations.

One interesting idea that has been proposed is that NGC 1851 is the result of a merger of two GCs \citep{Catelan1997}.  \citet{Carretta2010} support this scenario, positing two GCs with slightly different [Fe/H] and $\alpha$ element abundances and possibly up to a 1.5 Gyr age difference. Further support is provided by the quadrimodal distribution of CN molecular line strengths of RGB and AGB stars found by \citet{Campbell2012} which could be the result of the superposition of two "normal" bimodal populations.

All of this paints a picture of an object with a complicated chemical history. NGC 1851, like a handful of other GCs ($\omega$ Cen, M22, M2), has some evidence for an overall metallicity spread and contains an unusual stellar population enhanced by a factor of 2--3 in \cno, and also enhanced in s-process elements.

One of the current aims of globular cluster research is to identify the sources of these abundance variations, and the nucleosynthetic sites that contributed to the chemical enrichment of the population(s) with abundances that do not resemble field stars. The basic picture of globular cluster formation is that an initial primordial population formed with abundances similar to those of field stars. Some astrophysical objects provided the site for the required nuclear reactions, and were then able bring this chemically modified material to their surface where it could be ejected and `contaminate' the remaining intra-cluster gas, from which subsequent generations of stars formed. The main candidates for these astrophysical objects are intermediate mass AGB stars \citep{Ventura2008}, rotating AGB stars \citep{Decressin2009}, and fast-rotating massive stars \citep{Decressin2007}. For the clusters showing iron-peak variations, it could also be necessary to invoke Type II supernov\ae.

Overall the picture of cluster formation is not well understood. One method for constraining the proposed formation models is to probe the chemistry of stars. Certain elements and combinations of elements will show star-to-star variations and trends that allow for the elimination of specific models. One such constraint is the \cno\ sum. Models of fast-rotating massive stars indicate that this sum should be constant, while AGB stars should cause an increase (though it is possible to create models of AGB stars where the sum is constant). One of the aims of this paper is to determine \cno\ for a large sample of RGB stars in this cluster.

Another aim of this paper is to investigate the multiple populations of stars in NGC~1851 through the combination of spectral indices and elemental abundances derived from diatomic molecular features. Spectral indices are a well-used method for investigating stellar chemistry \citep[e.g.,][]{Norris1981,Beers1999,Morrison2000} as they can be used to quickly classify stars, and they rely on spectra with lower resolution than would be traditionally required for abundance analysis using equivalent widths. They are also very useful for quantifying broad molecular features. Although diatomic molecular features are more complicated in their spectra than the spectral lines formed from single elements, they are readily visible in low-resolution spectra which can be acquired with adequate signal in a reasonable amount of telescope time.

Finally we aim to investigate the astration of carbon during the giant branch phase of stellar evolution. Observations have shown the abundance of carbon decreases as the stars ascend the giant branch, and this has been seen in both globular cluster stars \citep[e.g.,][]{Martell2008a} and field stars \citep[e.g.,][]{Gratton2000,Smith2003}. These observations do not agree with the predictions of canonical models that there should be no mixing in low mass giant stars after they pass the first dredge-up event. Non-convective mixing must be taking place that is able to transport CN-rich material to the surface. This material must also reduce the $^{12}$C abundance. The exact mechanism for this non-convective deep mixing is not clear.

It is assumed that this deep mixing occurs when the abundance discontinuity created by the first dredge-up is removed by the outward progress of the hydrogen-burning shell. This discontinuity takes the form of a large gradient in mean molecular weight. We direct the reader to \cite{Karakas2014} for a review of the different proposed mechanisms for deep mixing. One possibility is the thermohaline mixing process \citep{Angelou2011,Angelou2012}, which could account for the observed abundance trends in some intermediate metallicity clusters. Another possibility is some sort of rotation-induced mechanism, although \cite{Palacios2006} and others have found that rotation models do not recreate the observed changes in the $^{12}$C/$^{13}$C ratio.

This paper has the following outline: Section \ref{sec:observations} is a description of the observations and their reduction; Section \ref{sec:spec_syn} describes how the C and N elemental abundances were determined; Section \ref{sec:spec_indices} describes the measurement of the spectral indices and comparisons with other studies; Section \ref{sec:newmembers} confirms the membership of potential new members found by \cite{Navin2015}; and Section \ref{sec:populations} defines four populations of stars in the cluster and how these populations are distributed spatially and chemically, in particular with regard to deep mixing and \cno.

\section{Observations and data reduction}\label{sec:observations}

\begin{table*}
\caption{Astrometric, photometric, and observational information for all stars. Positions and J \& K$_S$ photometry are from 2MASS; radial velocities from \citet{Navin2015}; V magnitudes are from \citet{Carretta2011}.}
  \label{table:basic_star_data}
  \begin{tabular}{ccccccccc}
  \hline
Star ID & RA(\degr) & Dec(\degr) & $v_r$ (km/s) & V&  J & K$_S$ & Exposure time (s) \\
  \hline
W05163977-4039434 & 79.16570804 & -40.66205269 & 288.0 &  & 13.429 & 13.164 & 2700\\
W05182129-4008551 & 79.58874971 & -40.14865385 & 292.4 &  & 14.205 & 13.646 & 5400\\
W05141567-4001444 & 78.56533307 & -40.02899735 & 301.1 &  & 14.483 & 13.783 & 5400\\
C26532 & 78.54720812 & -40.06923044 & 308.0 & 16.472 & 14.81 & 14.301 & 5400\\
C28913 & 78.50083292 & -40.06101996 & 308.1 & 15.633 & 13.758 & 13.149 & 5400\\
C25497 & 78.49679128 & -40.07338439 & 308.5 & 16.608 & 14.867 & 14.205 & 8100\\
C20827 & 78.58483312 & -40.10576223 & 310.0 & 17.012 & 15.506 & 14.632 & 2700\\
W05142069-3958560 & 78.58624947 & -39.9822268 & 311.2 & 15.628 & 14.218 & 13.651 & 5400\\
C40615 & 78.51841642 & -40.02385219 & 311.7 & 15.156 & 13.179 & 12.484 & 2700\\
C14827 & 78.42441639 & -40.07823161 & 312.1 & 16.994 & 15.342 & 14.926 & 8100\\
   \hline
  \end{tabular}

(This table is available in its entirety in a machine-readable form.)
\end{table*}

\begin{figure}
	\includegraphics[width=\columnwidth]{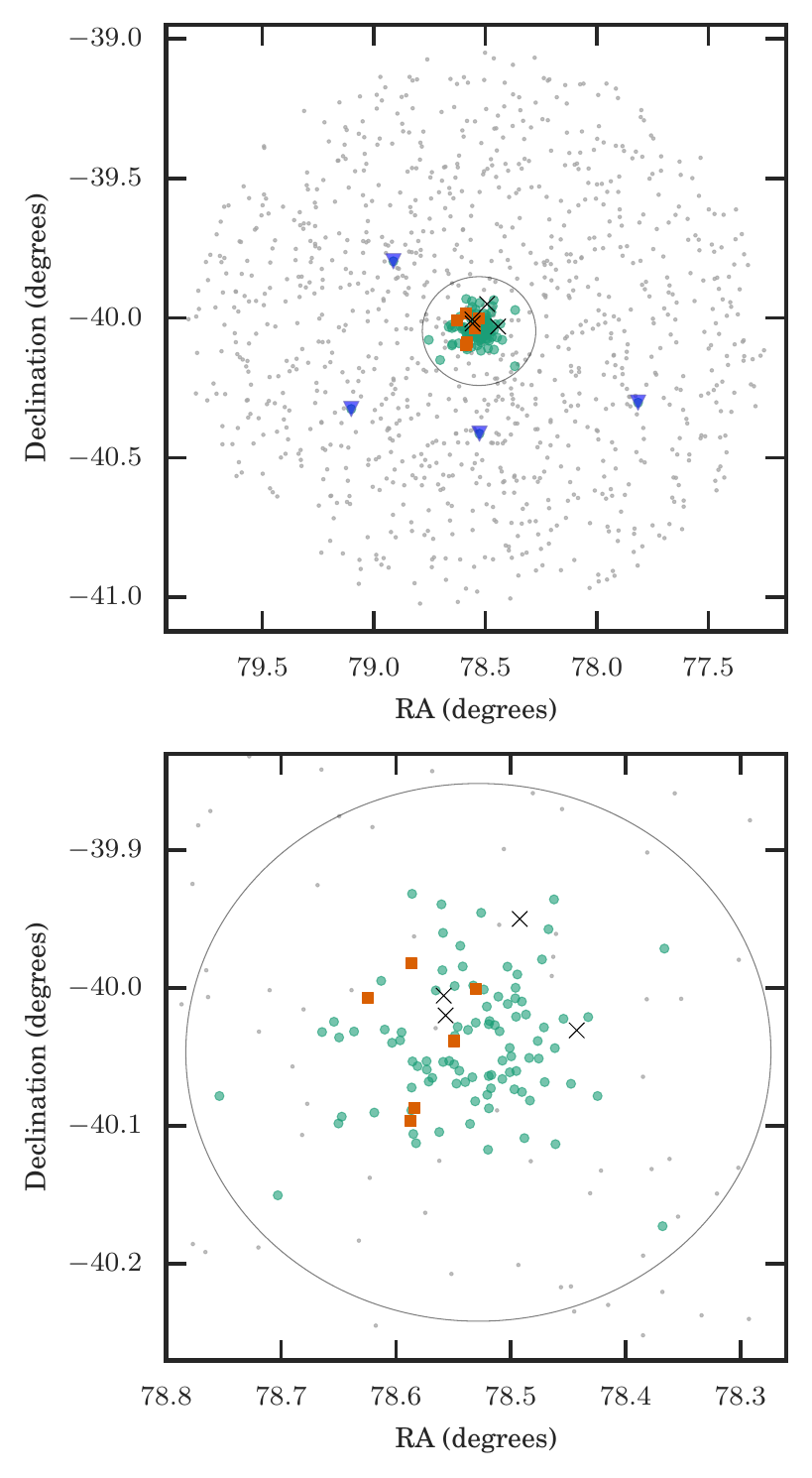}
    \caption{Sky position of the NGC~1851 stars observed. The bottom plot shows just the cluster region of the top plot. The circle is the 11.7~arcmin tidal radius of the cluster. Small grey dots are observed field stars \citep[based upon their radial velocities;][]{Navin2015}, large green circles are for previously known RGB cluster members from \citet{Carretta2011} and the other symbols are for new members found by \citet{Navin2015}: purple triangles are extratidal RGB members, orange squares are tidal giant branch members, and black crosses are four AGB tidal members with very weak spectral features (see Section \ref{sec:newmembers} for further discussion).}
    \label{fig:sky_plot}
\end{figure}

\begin{figure}
	\includegraphics[width=\columnwidth]{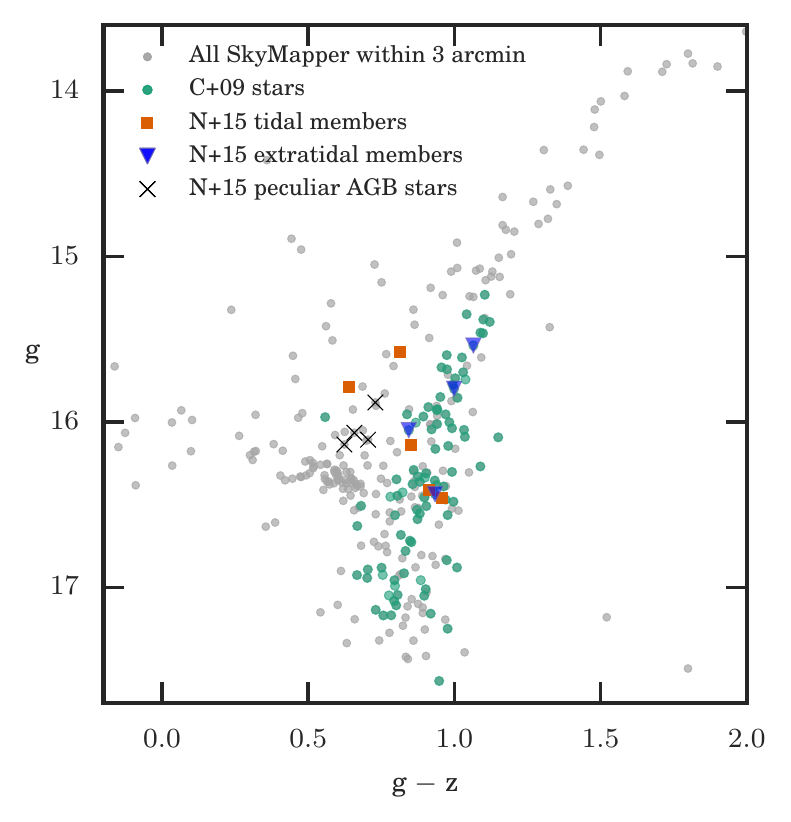}
	\caption{Colour-magnitude diagram for the stars observed. The photometry is from the Early Data Release (EDR) of SkyMapper \citep{SkyMapper2016}.}
    \label{fig:cmd}
\end{figure}

Spectra of 1119 stars in and around the globular cluster NGC~1851 were acquired on the nights of the 17 \& 18 December 2012 using the 3.9-metre Anglo-Australian Telescope and the multi-object, fibre-fed AAOmega spectrograph as part of program AAO2012B/18. A sky plot and a colour-magnitude diagram (CMD) of the stars observed are shown in Figures \ref{fig:sky_plot} and \ref{fig:cmd} respectively. Observational, astrometric and photometric data are given in Table \ref{table:basic_star_data}.

AAOmega is a medium resolution spectrograph, utilizing the 400-fibre 2dF fibre positioner, with which two wavelength regions can be acquired simultaneously \citep{Sharp2006}. In this paper, we concentrate on the spectra acquired using the blue 580V grating. This grating provides spectra with a spectral resolution of $R\sim$1200 for the wavelength range 3700--5800~\AA. The red spectra were acquired using the 1700D grating, which is designed for observing the near-infrared calcium triplet for determination of radial velocities. It had a wavelength coverage of 8340--8842~\AA\ with a spectral resolution of $R\sim$10000.

Three 2dF configurations were observed, centred on the cluster (RA = $01^{\rm h} 23^{\rm m} 42.77^{\rm s}$; Dec = $-14\degr 11\arcmin 49.8\arcsec$). The primary aim was to observe 95 cluster members that had previously been observed by \cite{Carretta2010}. The remainder of the fibres were allocated to stars from the 2MASS catalogue \citep{Skrutskie2006} within 1\degr\ of the cluster centre with the aim of identifying from their radial velocities cluster members with large angular separation from the cluster centre. \cite{Navin2015} were able to identify 13 previously unidentified cluster members from these 2MASS stars, of which four were located outside the tidal radius (Section \ref{sec:newmembers}).

It was planned for each star to have three exposures (3x2700~s) resulting in a signal-to-noise ratio per resolution element $>100$ for the faintest stars. Unfortunately, due to time constraints 23 per cent of stars (255/1119) received only one 2700s exposure, and 33 per cent (367/1119) received only 2x2700s. Two-thirds of stars have spectra with signal-to-noise $>100$ per resolution element.

\begin{figure}
	\includegraphics[width=\columnwidth]{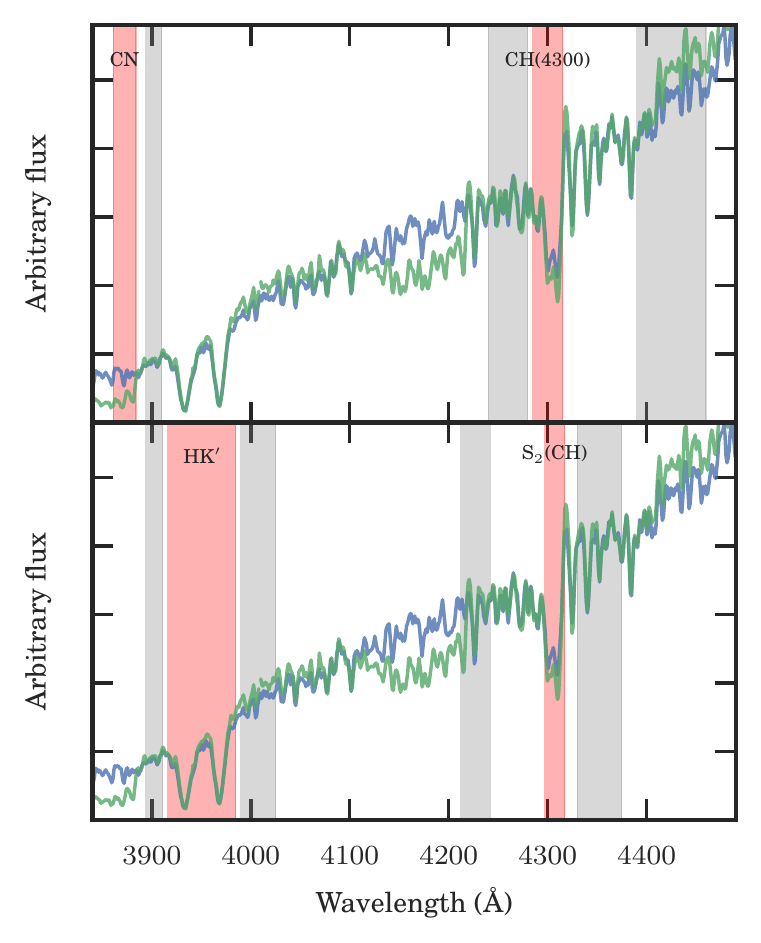}
    \caption{Both panels: Example spectra of two stars which have the same temperature ($\sim4700$ K) but differing molecular band strengths, C47795 (green spectrum; weaker CH and CN bands) and C20922 (blue spectrum; stronger CH and CN bands). Indicated in both panels are the regions used for the spectral indices. For each spectral index, the closest regions in grey were used for the continua. Top panel: the \cnblue\ CN spectral index and the \chold\ index in red. Bottom panel: the \hkdash\ and \sch\ indices in red.} 
    \label{fig:spectral_bands}
\end{figure}

The raw spectra were reduced using the AAO's {\sc 2dfdr} (v6.14) \citep{AAOSoftwareTeam2015}. We utilized {\sc 2dfdr}'s implementation of PyCosmic \citep{Husemann2012} to remove cosmic rays from the raw images. The raw images were bias-subtracted using the nightly master bias frame. Apart from these changes, the individual spectra were extracted and wavelength calibrated using the default settings provided for the 580V grating. We found that some of the line strengths were exaggerated by the  combining algorithm of \textsc{2dfdr}, so we co-added individual spectra after the \textsc{2dfdr} reduction. A spline was fitted to each individual spectrum of a star to give a `normalized' spectrum, then the average flux value at each wavelength point was determined, then one of the original splines was used to `de-normalize' the co-added spectra. Examples of the final spectral product are shown in Figure \ref{fig:spectral_bands}.

We used of radial velocities from \cite{Navin2015} to correct our reduced spectra to the rest wavelength. Cluster membership was assumed if the radial velocity was between 307 and 329 km/s. They showed that it was very unlikely for any field star to have such a radial velocity in that direction of the Galaxy.

\section{C and N abundance determination}\label{sec:spec_syn}
\begin{table*}
\caption{Spectral indices and elemental abundances for all stars. The spectral index definitions are given in Section \ref{sec:spec_indices}. The full version of the table also contains the uncertainties.}
  \label{table:spectral_indices}
  \begin{tabular}{cccccccccccccccccccc}
\hline
Star ID & \teff & \logg & \feh & \hkdash & \chold & \sch  & \cnblue & \ofe & \mgfe & \sife & \bafe & \cfe & \nfe \\
\hline
C13618 & 4590 & 1.95 & -$1.12$ & 0.54 & 1.07 & 1.74 & 0.18 & -$0.43$ & 0.31 & 0.42 & 0.48 & -$0.66$ & 1.37\\
C14827 & 4918 & 2.59 & -$1.17$ & 0.46 & 1.03 & 1.70 & 0.22 &  & 0.32 & 0.35 & 0.83 &  & \\
C16120 & 4721 & 2.19 & -$1.10$ & 0.55 & 1.04 & 1.72 & 0.15 & 0.24 & 0.35 & 0.35 & 0.70 & -$0.55$ & 1.47\\
C16984 & 4769 & 2.29 & -$1.13$ & 0.51 & 1.12 & 1.81 & -$0.29$ & 0.05 & 0.36 & 0.41 & 0.09 & -$0.21$ & 0.05\\
C20189 & 4874 & 2.54 & -$1.16$ & 0.45 & 1.03 & 1.72 & 0.01 & -$0.05$ & 0.39 & 0.40 & 0.36 & -$0.40$ & 1.60\\
C20399 & 4746 & 2.26 & -$1.14$ & 0.49 & 1.07 & 1.75 & 0.08 & -$0.05$ & 0.38 & 0.38 & 0.36 & -$0.48$ & 1.47\\
C20426 & 4839 & 2.46 & -$1.11$ & 0.47 & 1.09 & 1.79 & -$0.27$ & 0.14 & 0.42 & 0.39 & 0.29 & -$0.30$ & 0.28\\
C20653 & 4731 & 2.22 & -$1.18$ & 0.50 & 1.09 & 1.77 & 0.21 & 0.13 & 0.39 & 0.37 & 0.40 & -$0.33$ & 1.33\\
C20827 & 4923 & 2.60 & -$1.14$ & 0.43 & 1.06 & 1.72 & -$0.36$ & 0.36 & 0.46 & 0.36 & 0.42 & -$0.27$ & -$0.08$\\
C20922 & 4659 & 2.08 & -$1.14$ & 0.55 & 1.13 & 1.79 & -$0.17$ & 0.00 & 0.43 & 0.40 & 0.49 & -$0.32$ & 0.40\\
\hline
\end{tabular}

(This table is available in its entirety in a machine-readable form.)
\end{table*}
The primary aim of these observations was to determine \cfe\ and \nfe\ for a subset of stars that \cite{Carretta2010} had previously observed, and for which they had determined the stellar parameters and \ofe\ abundance. Of the 1119 stars observed for this work, 95 were observed by \cite{Carretta2010}, of which 71 had \ofe\ abundances (or upper limits) determined. We have adopted their previously derived stellar parameters and \ofe\ for these stars for our work.

In order to determine \cfe\ and \nfe\ from molecular features of CH and CN, it is necessary to know the oxygen abundance of the stars. This is because of the equilibria that exist between CO, CH, and CN in the stellar atmosphere, with CO strongly favoured in its formation. Therefore the \cfe\ determined from CH will depend on the \ofe\ adopted, and the \nfe\ from CN will depend on both \cfe\ and \ofe. The uncertainty in \ofe\ is the dominant term in the uncertainty for \nfe.

Our method for determining \cfe\ and \nfe\ made use of spectrum synthesis and spectral matching. For each star two spectral regions (3800--3915 \AA\ and 4100--4500 \AA) were synthesized for a wide range of possible \cfe\ and \nfe: $-1.5 < \cfe < 0.3$ and $-0.5 < \nfe < 1.9$ with 0.05 dex step sizes. These two regions cover three important CH- and CN-rich regions of spectrum in giant stars: the carbon-dependent \textit{G}-band region (mainly CH: $A^2\Delta - X^2\Pi$; 4295--4325 \AA) and two portions of the CN blue system ($B^2\Sigma - X^2\Sigma$; 3861--3884 \AA\ \& 4195--4222 \AA). This gave a total of 1900 synthetic spectra to compare with each observed spectrum in each region.

The spectrum synthesis used a modified version of the local thermodynamic equilibrium spectrum synthesis software \textsc{MOOG} \citep{Sneden1973}\footnote{The modified version, created by Ryan T. Hamilton, produces output in a FITS format: \url{https://bitbucket.org/astrobokonon/moogifications}.}. \cite{Kirby2011} interpolated a grid of Castelli-Kurucz models which we used for this work. The model with the closest matching stellar parameters (\teff, \logg, \feh) was used for each star. For giant stars the grid had models every 100~K in temperature, with 0.5~dex steps in \logg\ and \feh\ (the $\feh = -1.0$ models were used). \cite{Kirby2011} provided appropriate values for the microturbulence for stars of those temperatures and gravities.

The changing $^{12}\textrm{C} / ^{13}\textrm{C}$ ratio of stars as they ascend the giant branch (the deep mixing process) was taken into account. The Sun has a ratio of about 80, while at the top of the giant branch it is closer to 6. We used the formulation from \cite{Kirby2015}, which they derived from figure 4 in \cite{Keller2001},
\begin{equation}
 ^{12}\textrm{C} / ^{13}\textrm{C} =
  \begin{cases}
   50                     & \text{if } \log g > 2.7\\
   63\times \log g - 120  & \text{if } 2.0 < \log g \leq 2.7 \\
   6                      & \text{if } \log g \leq 2.0
  \end{cases}
\end{equation}

For the 3800--3915 \AA\ region a line list from \textsc{spectrum} was used \citep{Gray1994}. For 4100--4500 \AA\ the line list created by \cite{Kirby2015} was adopted.

The method used for determining the \cfe\ and \nfe\ abundances was a variation on the method used by \cite{Kirby2015}. The first region of the observed spectrum that was analyzed was from 4100--4500 \AA\, in order to derive the \cfe\ abundance from 4295--4325 \AA. An initial flux normalization was performed for the whole observed spectrum (3700--5800 \AA) by fitting a fifth-degree univariate spline with knots every 250 \AA. This was iteratively fitted to the spectrum, with spectrum points rejected if they were more than 3 standard deviations above or 0.1 standard deviations below the spline.

We initially assumed that $\nfe = 1.0$. The nitrogen abundance does not have a large effect on \cfe\ across the nominal range of nitrogen for cluster stars. For each synthetic spectrum with $\nfe = 1.0$ the following procedure was used. The flux normalized observed spectrum was divided by the synthetic spectrum. A spline was fitted through this quotient, and this spline was used to correct the initial normalization for the actual continuum level of that synthetic spectrum. The $\chi^2$ was then found for the wavelength ranges 4260--4325 \AA.

From the full range of \cfe\ in the spectrum grid, the five smallest $\chi^2$ values were found and parabola fitted to determine the best \cfe. This \cfe\ value was then used to determine \nfe\ from 3861--3884 \AA. We iterated this procedure until it converged.

Errors for the abundances were determined by varying each stellar parameter by its own uncertainty, and then adding in quadrature the change in abundances this caused for \cfe\ and \nfe\ (we note that this ignores any co-variance that might exist). Although the grid was relatively coarse it was found that varying that varying $\logg$ by one grid point would typically affect \cfe\ by $<0.03$~dex. As would be expected, changing the metallicity grid point by $0.1$~dex had the opposite affect on the \cfe, while the affect of changing the gravity by 100~K was of a similar magnitude. For nitrogen, the effect of moving one model grid point was larger, typically 0.2~dex for any grid point change of the \teff, \logg, or \feh. In both cases the \feh\ was the largest source of the error for the abundances.

The average errors for \cfe\ and \nfe\ are $\pm0.11$ and $\pm0.48$ dex respectively. The relatively large uncertainty for \nfe\ is caused by a number of factors. First, we are using a region of the spectrum that has low signal due to the spectral energy distribution of the stars. As shown in Figure \ref{fig:spectral_bands}, the CN region has 3--4 times less flux than the CH region. Second, the determination of the \nfe\ from CN depends on knowledge of \cfe\ and \ofe, both of which cause additional uncertainty. With this in mind, we will be restrained in our conclusions for \nfe\ and \nfe-derived quantities.

Our \cfe\ and \nfe\ abundances are given in Table \ref{table:spectral_indices}. We defer comparisons with other studies to the end of the next section.

\section{Spectral indices}\label{sec:spec_indices}
Spectral indices are a measure of the strength of a given spectral feature with respect to surrounding continuum regions. As discussed in Section \ref{sec:spec_syn}, the determination of the \cfe\ and \nfe\ abundances requires knowledge of the \ofe\ abundance, which itself requires high resolution spectra of the stars. Spectral indices do not have this requirement but do need to be considered with respect to other known stellar parameters of the star.

There are a variety of definitions of spectral indices, especially for the CN- and CH-rich regions of the spectrum. For this work, four different indices were measured: a CN index, two different indices for the CH 4300\AA\ band, and a Ca H \& K index. They are shown in Figure \ref{fig:spectral_bands} along with examples of two stars with the same temperatures but different CN and CH band strengths. Our indices result are given in Table \ref{table:spectral_indices}.

We define that for the spectral flux as a function of wavelength:
\begin{equation}
F_{A-B} = \int_{A}^{B}F(\lambda) d\lambda.
\end{equation}

We measured a spectral index for the CN around 3883\AA, as defined by \cite{Harbeck2003},
\begin{equation}
\cnblue = -2.5 \log \frac{F_{3861 - 3884}}{F_{3894-3910}}.
\end{equation}

For CH we used two definitions; one from \cite{Martell2008},
\begin{equation}
\sch = -2.5 \log \frac{F_{4297 - 4317}}{F_{4212-4242} + F_{4330-4375}},
\end{equation}
and one from \cite{Harbeck2003},
\begin{equation}
\chold = -2.5 \log \frac{F_{4285 - 4315}}{0.5F_{4240-4280} + 0.5F_{4390-4460}}.
\end{equation}
We primarily used the \sch\ index, but also measured the \chold\ index for comparison with other studies.

\cite{Lim2015} presented a new calcium H \& K spectral index,
\begin{equation}
\hkdash = -2.5 \log \frac{F_{3916 - 3985}}{2F_{3894-3911} + F_{3990-4025}}.
\end{equation}
Although the calcium strength was not the primary aim of this research, \cite{Lim2015} included an analysis of NGC~1851 stars that were in common with our sample, so we measured \hkdash\ in order for us to allow a comparison of the results from the two studies. \hkdash\ provides a useful metallicity indicator and a discriminator between horizontal and giant branch stars. It is also often used as a cluster/field star discriminant.

\begin{figure}
	\includegraphics[width=\columnwidth]{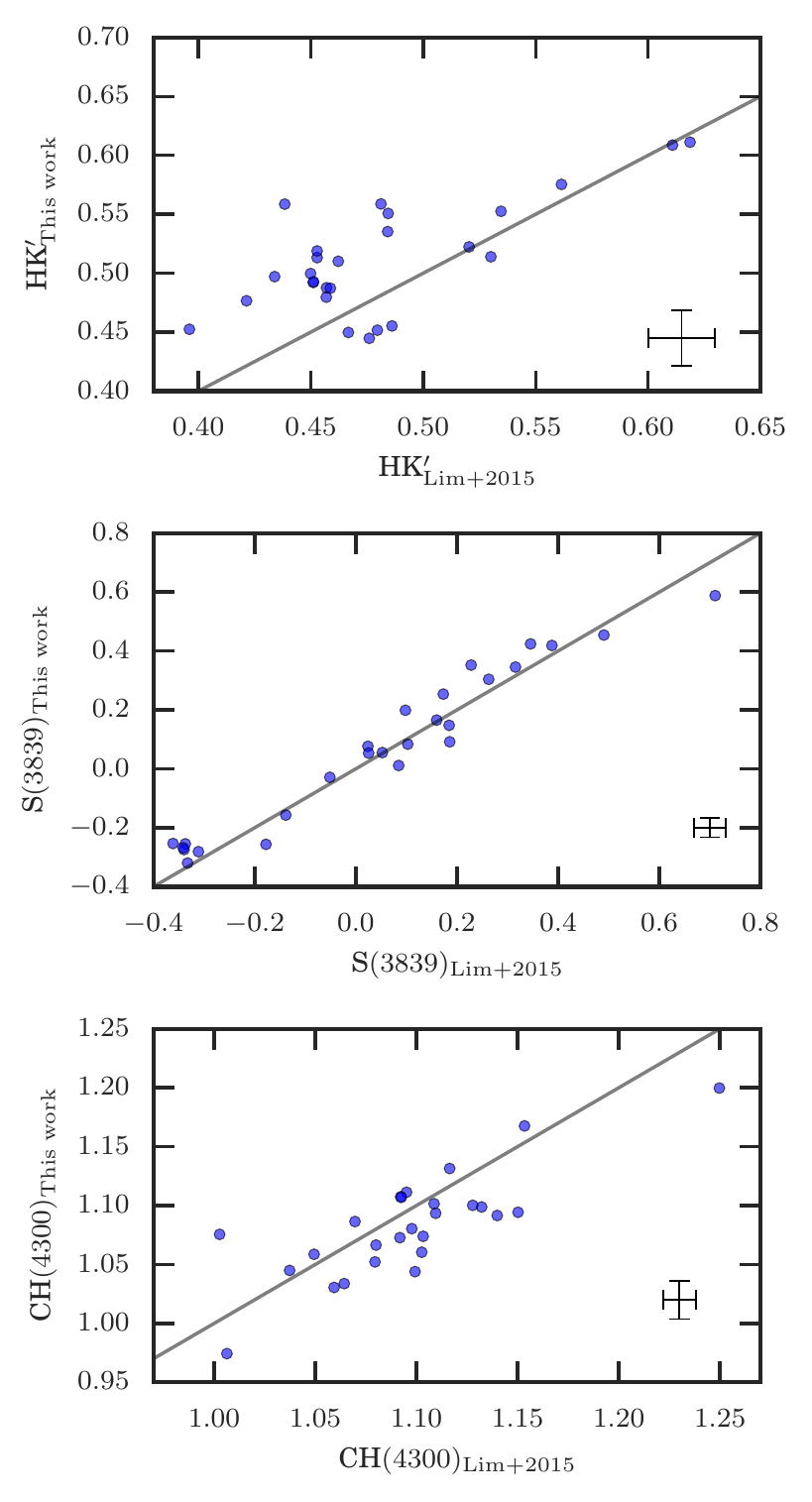}
    \caption{Comparison of the spectral index strengths for the 28 stars in common with \citet{Lim2015} and this work. The straight line is a one-to-one line. Also shown are the average uncertainties of the spectral indices.}
    \label{fig:hkdash_compare}
\end{figure}

We had 28 stars in common with \cite{Lim2015}. There is a general correlation between the spectral indices derived by the two studies (Figure \ref{fig:hkdash_compare}). The range of \hkdash\ is much smaller than that of \cnblue\ which magnifies the scatter in the comparison. The $\langle\Delta \hkdash \rangle = -0.03\pm0.04$ and $\langle\Delta \cnblue \rangle= -0.02\pm0.06$.

\begin{figure}
	\includegraphics[width=\columnwidth]{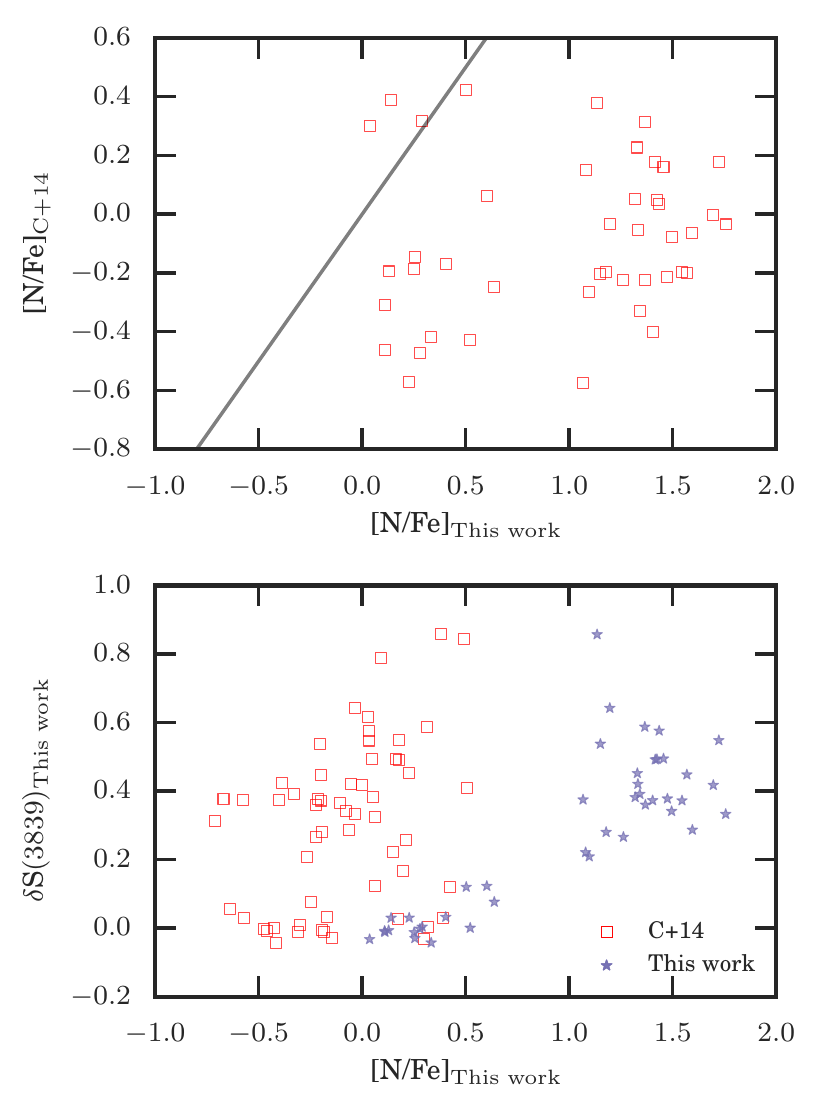}
    \caption{Top: Comparison of the \nfe\ abundances determined by this study and \citet{Carretta2014a}. The line is the one-to-one line. Bottom: \nfe\ abundances from this work and \citet{Carretta2014a} compared to the \dcnblue\ spectral index strength determined from our spectra. It would be expected that there would some correlation between the \nfe\ and \dcnblue\ strength of stars.}
    \label{fig:C14_comparison}
\end{figure}

For comparison of \cfe\ and \nfe, there was only one large study in common with ours, which was performed by \cite{Carretta2014a}. There were 60 stars in common for which \nfe\ was determined. These were stars that had previously been analyzed in \cite{Carretta2011}, hence the large crossover in our two studies. Unfortunately, there is little correlation between the \nfe\ values determined by the our two studies, nor with their \nfe\ and our \cnblue\ (Figure \ref{fig:C14_comparison}). The fact that there is little correlation between nitrogen abundances of \cite{Carretta2014a} and our spectral index for the blue CN feature suggests there may be a problem with their \nfe\ estimates. They assumed that $\cfe = 0.0$ for all stars, which would be incorrect in light of the molecular equilibria that exist between CO, CH, and CN (Section \ref{sec:spec_syn}). The spectra they used were acquired for the Gaia-ESO Survey, and so do not have an optimal signal-to-noise for the task of abundance analysis for elements like nitrogen.

Both observationally and theoretically we would expect that CN strength and nitrogen abundance should track each other. We note that their \nfe\ values appear to have no correlation with their previously determined \nafe\ for the same stars, when there should be a strong correlation between sodium and nitrogen abundances in globular cluster giant stars \citep[e.g.,][]{Norris1985,Marino2012,Smith2013}.

\section{Confirmation of new members}\label{sec:newmembers}
\begin{figure}
	\includegraphics[width=\columnwidth]{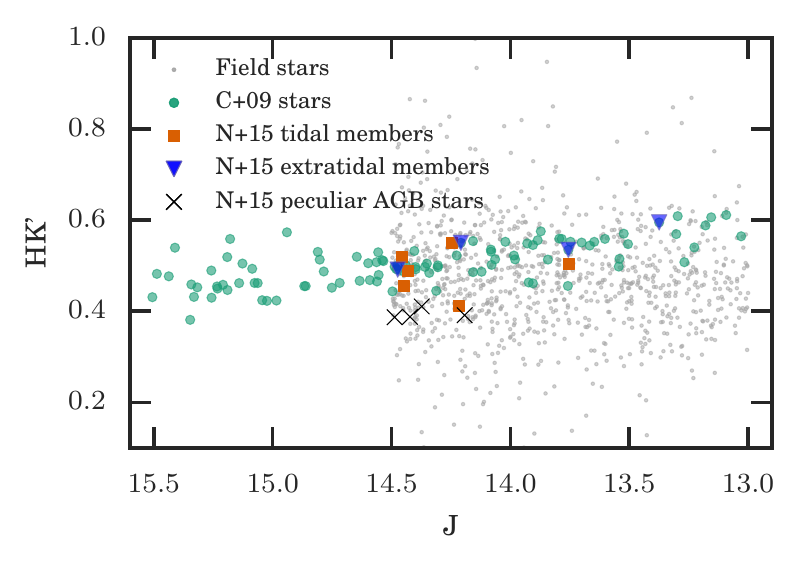}
    \caption{The \hkdash\ indices determined for the sample of stars in NGC 1851. The symbols are as defined in Section \ref{sec:newmembers} and Figure \ref{fig:sky_plot}.}
    \label{fig:j_hk}
\end{figure}

\begin{figure}
	\includegraphics[width=\columnwidth]{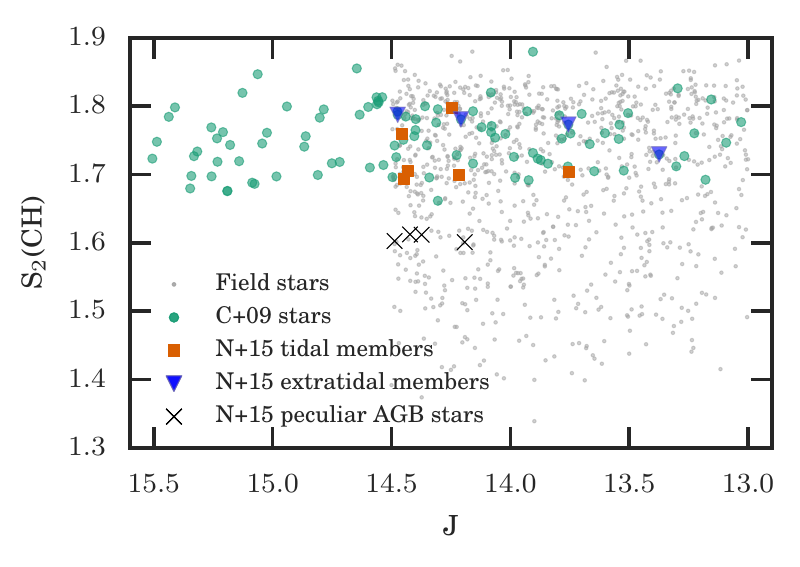}
    \caption{The dependence of \sch\ on the magnitude of the star. The symbols are as defined in Section \ref{sec:newmembers} and Figure \ref{fig:sky_plot}. The four new peculiar AGB members are clear outliers from the rest of the cluster stars.}
    \label{fig:j_S2CH}
\end{figure}

\begin{figure}
	\includegraphics[width=\columnwidth]{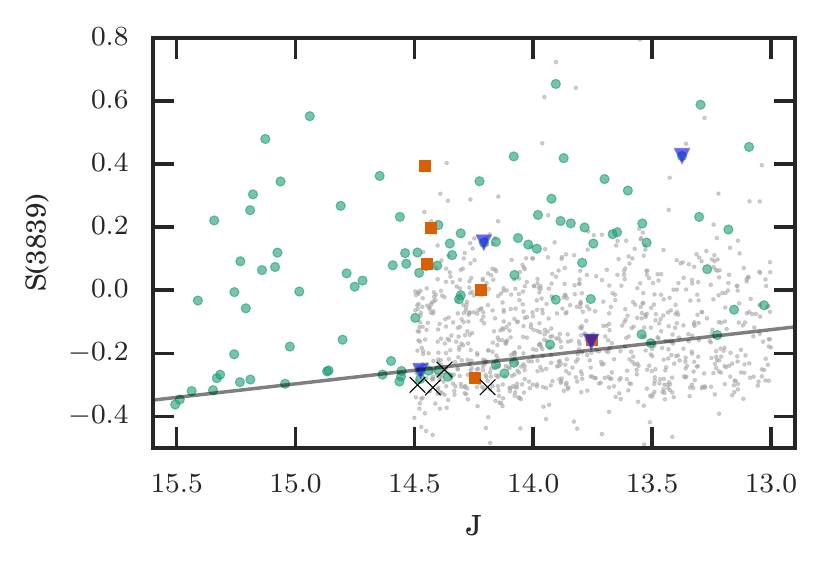}
	\includegraphics[width=\columnwidth]{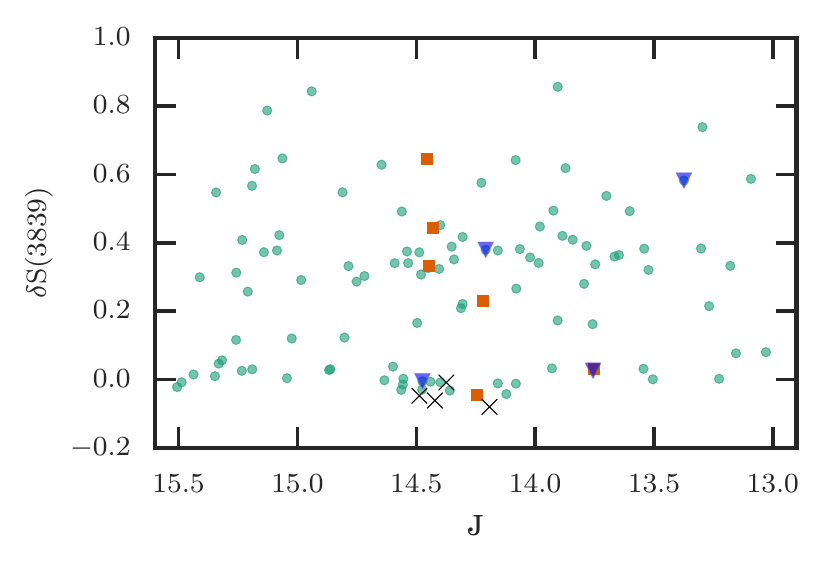}
    \caption{The \cnblue\ index results. The symbols are as defined in Section \ref{sec:newmembers} and Figure \ref{fig:sky_plot}. One of the extratidal stars is coincident on this plot with a tidal star. Top: The dependence of \cnblue\ on the magnitude of the star. The black line represents an empirical minimum value of the CN index at a given J magnitude. This line is used to convert \cnblue\ to \dcnblue. Bottom: The temperature correction applied ($\dcnblue = \cnblue - [ -0.078\times J+0.874]$). We have not plotted the field stars as the temperature correction cannot be applied to them.}
    \label{fig:j_S_CN3839}
\end{figure}

\cite{Navin2015} identified 13 new members of NGC~1851 using the radial velocities as determined from the calcium triplet spectral lines. A reanalysis of the \cite{Navin2015} data set identified one further star. In this section we discuss these 14 stars and how their spectral indices compare to the rest of the known cluster members.

In Figures \ref{fig:j_hk}, \ref{fig:j_S2CH} and \ref{fig:j_S_CN3839} we present the spectral indices, both for the field stars (grey dots) and cluster members (large green dots). The new stars from \cite{Navin2015} are shown with different symbols, with the classifications justified in this Section. Their positions on the sky and on the CMD are shown in Figures \ref{fig:sky_plot} and \ref{fig:cmd} respectively.

There were six new member stars which were very close to the cluster centre, with angular distances less than 6~arcmin, placing them well inside the tidal radius \citep[11.7~arcmin;][]{Harris1996-mu}. Their spectral indices are all very similar to the known cluster members. In particular, four of the six are CN-enhanced, which is different from field stars (Figure \ref{fig:j_S_CN3839}). 

Another four new members identified have very weak \hkdash\ (Figure \ref{fig:j_hk}) and \sch\ (Figure \ref{fig:j_S2CH}) indices. As shown in Figure \ref{fig:cmd}, these four stars appear to be AGB stars. During the refereeing process of this paper the SkyMapper Early Data Release photometry \citep{Keller2007,SkyMapper2016} became available. This allowed us to determine the evolutionary stage of these stars. Previously we had thought they were HB stars for which the stellar atmosphere is very different from giant branch stars, which explained their peculiar index strengths. But since they are in fact AGB stars, it is unclear why these stars have such weak \hkdash\ and \sch\ indices compared to other giant branch stars. Figure \ref{fig:cmd} shows that seven AGB stars were observed. As well as these anomalously weak four stars, one of the other tidal members also has weak \hkdash, but not \sch. Further investigation of all the AGB stars is required \citep[especially in light of the results from][]{Campbell2012,MacLean2016}, in particular why some have such weak spectral indices, but that is beyond the scope of this paper. 

The four extratidal members identified all have angular distances larger than 22~arcmin, well outside the tidal radius. Again they show spectral index strengths consistent with the rest of the cluster. Three are CH-rich (Figure \ref{fig:j_S2CH}) and two of the stars are CN-rich (Figure \ref{fig:j_S_CN3839}). With only four stars it is difficult to conclude if there is any preference for these stars to be CH- and/or CN- enhanced. But at least for this very small sample of four stars, we find that there are some CN and CH enhanced stars, which would presumably come from the later generations of stars.

\cite{Kunder2014} used the RAVE survey to search for members of NGC~1851. They were able to identify 11 extratidal stars with velocities consistent with that of the cluster. Unfortunately only one of the stars had individual elemental abundance determined (in this case magnesium), so it is not possible to compare our studies for similarities in the chemistry of these escaped stars.

It is of interest to make a simple calculation to compare the destruction rate implied by the detection of four extratidal stars to that of simulations.

\citet{Gnedin1997} paper calculated the total destruction rate for an isotropic GC system distribution with a Weinberg adiabatic conservation factor for the shock processes. Their values for NGC 1851 of $1.27\times10^{-11}$\ and\ $1.24\times10^{-11}~yr^{-1}$\ for two different Galactic models are slightly lower than the median destruction rates for their sample of 119 GCs for the same setup. \citet{Moreno2014} calculated the total destruction rate due to bulge-bar and disk shocking. This ranged from $8.31\times10^{-16}$\ to\ $1.03\times10^{-15}~yr^{-1}$ depending on the models used. This is considerably lower than their median values and the values calculated by \citet{Gnedin1997}.

We set the observed cluster fractional mass loss as the ratio of the total cluster extratidal halo stars V luminosity to the integrated cluster V luminosity (from the cluster integrated V magnitude). The observed cluster fractional mass loss is 0.0022. We estimated sample completeness by comparing the number of unique observed stars outside the tidal radius (989) to the number of stars (1128) in the 2MASS catalog (quoted completeness of $J$=15.8 mag) that (i) are outside the tidal radius, and (ii) have the same photometric constraints as the third group of target stars ($0<J-K_s<1.2$ and $13<J<14.5$). This gave a sample completeness of 0.877 for our final cluster extratidal halo sample of four stars. The fractional mass loss was divided by the completeness to give a total fractional mass loss of 0.025.

We then estimated the time taken for a star to move outside our 1\degr\ field of view. As an estimate, equation 18 of \citet{Kupper2010} gives the relative velocity of escaped stars for clusters in circular orbits in the disc. For NGC 1851 this gives a relative velocity of $\pm\sim$5.9~km/s.  \=textbf{As stars can escape in any direction, the mean relative velocity perpendicular to our line-of-sight (i.e. the proper motion) is $\pm\sim$5.9$\times2/\pi =\ \pm\sim$3.8~km/s. The stars would move 1\degr\ from the NGC 1851 central position in $\sim$55 Myr at this velocity.

We divided the fractional mass loss by the time taken for the cluster extratidal halo stars to move outside our 1\degr\ field of view to give a cluster destruction rate of $4.6\times10^{-11}\ yr^{-1}$.

There are several assumptions and estimates in this calculation that should be mentioned: (i) The two samples probably have different mass-to-light ratios. The integrated cluster luminosity includes a contribution by dwarfs, whereas the observed total luminosity of the cluster extratidal halo stars does not, as they are too faint to detect in our selection criteria. Therefore the observed cluster fractional mass loss is likely underestimated; this would translate to an underestimate of the destruction rate. (ii) The estimate of the velocities of escaped stars is, as stated, strictly applicable to clusters in circular orbits in the disc; if the actual velocities are lower then the calculated destruction rates would be an overestimate and vice versa. (iii) The velocities of escaped stars are constant. Studies have been done of the variation of $V_r$ along the tidal tail of GCs (e.g \citealt{Odenkirchen2009} and \citealt{Kuzma2014} find gradients of 1.0 $\pm$ 0.1 km s$^{-1}$ deg$^{-1}$ for Palomar 5), but without a detailed model for the gravitational potential of the MW's dark matter halo to calculate the orbit of escaped stars, the effect on the destruction rate is difficult to estimate.

Given those limitations, our estimated destruction rate is comparable to the destruction rate calculated by \citet{Gnedin1997}, but considerably higher than that of \citet{Moreno2014}. The final number confirmed as ex-cluster members can constrain theoretical studies of GC destruction rates as well as the contribution of GCs to the Galaxy's stellar halo. However there are also significant differences in the predicted rates, so the models and simulations are also not yet definitive, and this may also account for some differences.

\section{Populations defined from CH and CN}\label{sec:populations}

\begin{figure}
	\includegraphics[width=\columnwidth]{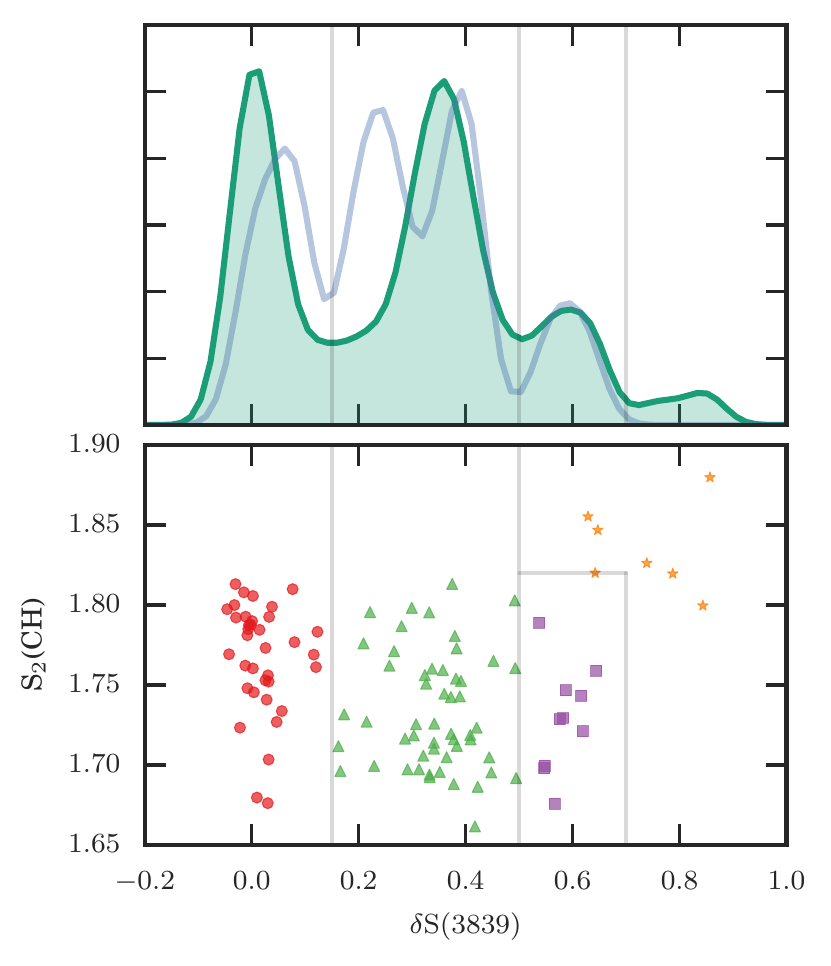}
    \caption{Top panel: Comparison of the distribution of \cnblue\ determined by this work (shaded green KDE with $\sigma = 0.15$) and \citet{Campbell2012} (light blue line; $\sigma = 0.2$). \citet{Campbell2012} suggested that there were four peaks. One of those peaks at about 0.25 appears to have been the result of small number statistics. It should be noted that the shared peak at 0 is just the result of the zeroing of the \cnblue. Bottom panel: For reference the \cnblue\ of the stars against the \sch\ of the stars, coded with the CN-CH populations.}
    \label{fig:j_dCN_kde_all}
\end{figure}

Inhomogeneity of CN and CH molecular band strengths in the spectra of giant branch stars was one of the first indications that the previously held paradigm that globular clusters are simple stellar populations was incorrect. In the case of NGC~1851, the fact that there are both CN-weak and CN-strong stars has been known since \cite{Hesser1982}. Recently it has emerged that NGC~1851 is not bimodal in its CN strength, like some other clusters, but exhibits multi-modal behaviour \citep{Campbell2012}.

From Figures \ref{fig:j_hk}, \ref{fig:j_S2CH} and \ref{fig:j_S_CN3839}, two features are clear: the background stars do not follow the same trends as the cluster members; and the cluster members exhibit multi-modal behaviour in their CN but not their CH.

The top panel of Figure \ref{fig:j_S_CN3839} shows the \cnblue\ index strengths for the cluster and non-cluster stars. The strength of spectral features depends not only on the abundance of that molecule in the stellar atmosphere, but also, to first order the temperature of the star: molecular features typically get weaker as the star gets hotter, which is especially true for CN. Consequently, it is necessary to correct for temperature to understand the distribution of the CN strengths in the cluster. Following the procedure from \cite{Norris1981}, we did this correction using J magnitude of the stars as a proxy for luminosity}. The bottom envelope was identified and a straight line fitted through this. The stars were then all corrected such that the envelope stars all have $\dcnblue\simeq0$.

We used a univariate kernel density estimate (KDE) to further investigate the multi-modal behaviour of \dcnblue\ (Figure \ref{fig:j_dCN_kde_all}). This showed that there were four peaks in this distribution. In the top panel of Figure \ref{fig:j_dCN_kde_all} we also show the distribution found by \cite{Campbell2012} for their sample of RGB stars which they found evidence for four peaks. However, their sample of 17 stars was much smaller than our sample, and it is likely that their peak at $\dcnblue \sim 0.25$ is the result of small number statistics, although we do note that there appears to be an excess of stars between the first two peaks of the KDE distribution.

Four populations of stars were defined from the combination of our \dcnblue\ and \sch\ spectral indices. We adopt the convention of other studies of multiple populations of globular clusters and refer to these as the primordial, intermediate, extreme, and anomalous populations. These population definitions are visually represented on the bottom panel of Figure \ref{fig:j_dCN_kde_all}:
\begin{itemize}
	\item primordial: $\dcnblue\leq0.15$ (fraction of total sample: 36/105; 34\%)
	\item intermediate: $0.16<\dcnblue\leq0.5$ (52/105; 50\%)
	\item extreme: $0.5>\dcnblue\leq0.7$ and $\sch<1.82$ (10/105; 9.5\%)
	\item anomalous: $\dcnblue>0.7$ or $\sch>1.82$ (7/105; 6.7\%)
\end{itemize}
We shall also refer to the primordial population as the first generation and the other three populations collectively as the second generation, though it is not clear if these populations are coeval or not. The boundary definitions were determined from the local minima in the KDE distribution. The index strengths and the temperature correction will have some uncertainty, so some stars near boundary edges could actually belong to adjacent populations.

For the whole sample of cluster stars, there is no obvious bimodality of the \sch\ index strength. However when the individual populations are considered, we find that the primordial population on average has slightly stronger CH bands than the intermediate and extreme populations.

\cite{Carretta2009} defined populations based upon the location of a star in the sodium-oxygen anticorrelation rather than its \dcnblue\ and \sch. Using these definitions, \cite{Carretta2011} found the following population sizes for NGC~1851: primordial, 32\%; intermediate, 66\%; extreme, 6.3\%.

In the following subsections we will discuss our CN-CH defined populations in the context of their elemental abundances (Section \ref{sec:elemental_abundances}), deep mixing (Section \ref{sec:deepmixing}), \cno\ (Section \ref{sec:cno}), and radial distribution (Section \ref{sec:radial}), as well as looking in particular the anomalous population (Section \ref{sec:anomalous}).

\subsection{Deep mixing}\label{sec:deepmixing}
\begin{figure}
	\includegraphics[width=\columnwidth]{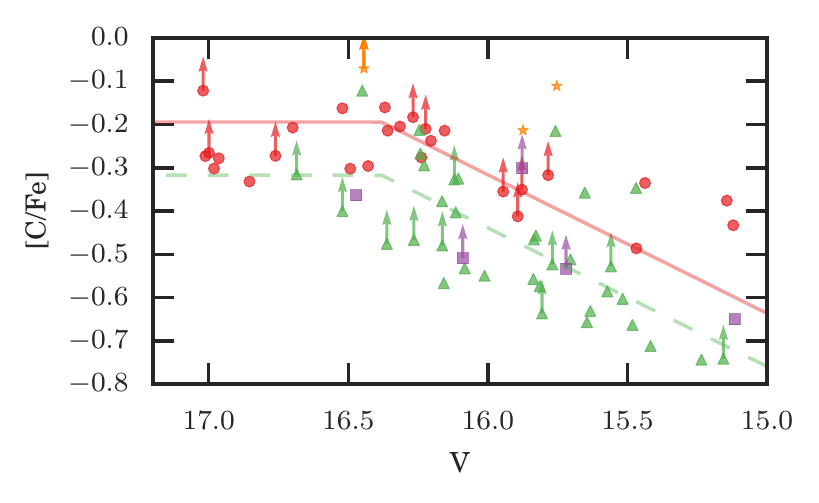}
    \caption{Change in \cfe\ with V magnitude. The symbols are as defined in Figure \ref{fig:j_dCN_kde_all}. The solid line is the best fit line for the primordial population, and the dashed line is for the intermediate population. Data points with arrows indicate those stars for which the carbon abundance is only a lower limit.}
    \label{fig:teff_cfe}
\end{figure}

As discussed in Section \ref{sec:introduction}, deep mixing entails the non-canonical depletion of the surface carbon abundance as stars ascend the giant branch. It is not expected to change the surface \cno\ abundance. We argue here that based upon their carbon astration, our CN-CH populations show different carbon abundances at a fixed luminosity, but with the same rate of depletion of surface carbon abundance.

With the carbon astration rate and pre-bump \cfe\ as free parameters, the best fit for the primordial population is $\cfe_{\textrm{pre-bump}}=-0.24$ and $\textrm{d}\cfe/\textrm{dV}=0.16$ dex per magnitude. For the intermediate population we find $\cfe_{\textrm{pre-bump}}=-0.32$ and $\textrm{d}\cfe/\textrm{dV}=0.32$ dex per magnitude.

There is no theoretical expectation, nor observational precedent, for stars of the same metallicity to experience different rates of deep mixing. So we next only allowed only the pre-bump \cfe\ to be the free parameter, and assumed that the carbon astration rate should be the same for both populations. For a cluster of the metallicity of NGC~1851, \cite{Martell2008a} predicts a carbon depletion rate of between 0.14 and 0.32 dex per magnitude. The models presented in \cite{Angelou2012} give a depletion of carbon of 0.33 dex per magnitude for clusters of similar metallicity to NGC~1851. A rate of 0.32 dex per magnitude provides a reasonable fit to the data, with $\cfe_{\textrm{pre-bump}}=-0.19$ and $-0.32$ for the primordial and intermediate populations. This change in slope shifts the primordial populations pre-bump \cfe\ only 0.05 dex carbon-richer, well within our uncertainties.

In Figure \ref{fig:teff_cfe} we show the \cfe\ abundances of stars as they ascend the giant branch (represented by their V magnitude). The stars are coded with their CN-CH populations. For the primordial and intermediate population we plot piecewise functions, with a constant carbon abundance below the luminosity bump and then a constant rate of depletion as the stars evolve past the bump. The luminosity bump magnitude was determined to be $V=16.38$ using \cite{Nataf2013} with $\feh = -1.18$, $[\alpha/\textrm{Fe}] = 0.3$, and $m-M = 15.47$.

It would be useful to acquire observations of fainter stars to constrain the pre-bump \cfe, especially for the intermediate population. For this population there are only three stars fainter than the bump, which makes it difficult to fit the pre-bump \cfe\ abundance. With so few stars, no attempt is made at fitting the extreme and anomalous populations. However there is a general downwards trend for the extreme population.

\subsection{Elemental abundances}\label{sec:elemental_abundances}
\begin{figure}
	\includegraphics[width=\columnwidth]{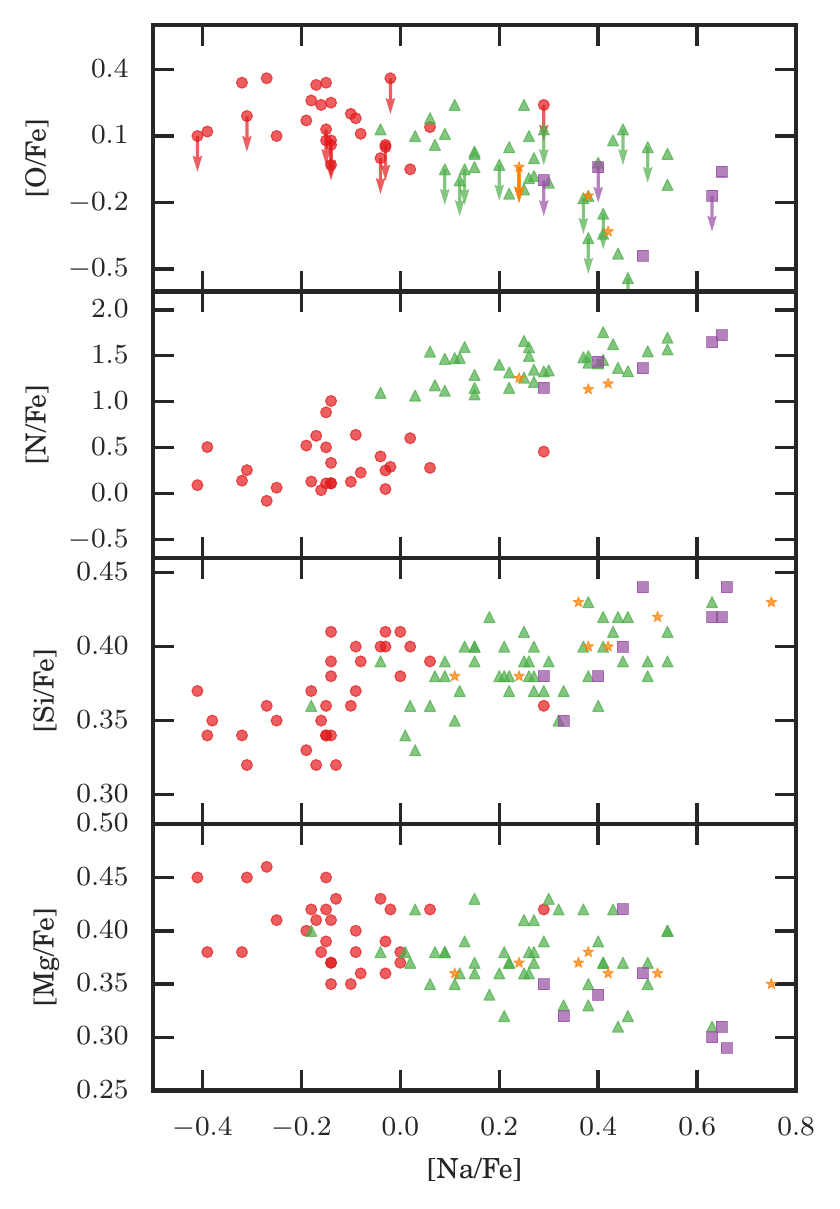}
	\caption{The correlations between sodium and oxygen, nitrogen and silicon and magnesium. The symbols are as defined in Figure \ref{fig:j_dCN_kde_all}. \nfe\ is from this work and the other abundances are from \citet{Carretta2011}.}
    \label{fig:chem_populations}
\end{figure}

\begin{figure}
	\includegraphics[width=\columnwidth]{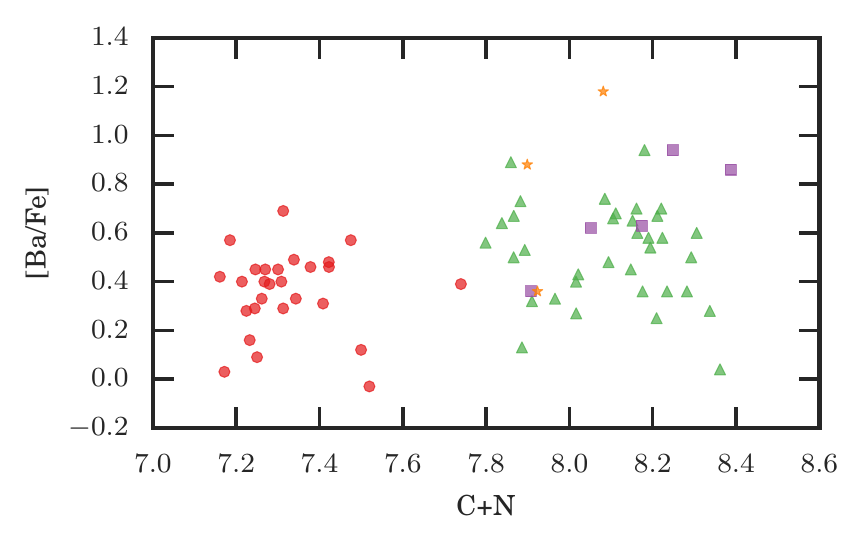}
    \caption{The correlations between the C+N and \bafe. The symbols are as defined in Figure \ref{fig:j_dCN_kde_all}. The elemental abundances are from \citet{Carretta2011}.}
    \label{fig:cn_bafe}
\end{figure}

There were 67 stars in common with \cite{Carretta2010} for which \cfe\ and \nfe\ could be determined. In Figure \ref{fig:chem_populations} we have combined our \nfe\ abundances with the elemental abundances from \cite{Carretta2010}. In addition we have coded each star with its CN-CH defined population.

As would be expected for populations defined primarily from their CN band strength, the first and second generation populations are clearly separated in their \nfe\ (Figure \ref{fig:chem_populations}). However, within the second generation, the intermediate, extreme, and anomalous populations are not not distinct. The distinction between the first and second generations is also present in their sodium abundances, but it is not so clear in oxygen and carbon. 

One of the previously identified features of NGC~1851 is that it exhibits a range of s-process elemental abundances. We chose barium as our s-process element as it was also measured successfully by \cite{Carretta2011} for the majority of stars. In Figure \ref{fig:cn_bafe} the carbon and nitrogen abundances are combined, which shows there is a clear separation between the first and second generation stars. Again, however, the CN-richest populations are not clearly distinct from the intermediate population. There is a slight trend towards increasing barium with increasing CN. This implies that those elements are produced by the same source, or in different sources with similar timescales.

\cite{Marino2014} identified 23 stars in the ``halo'' around NGC~1851 at much larger distances than the stars found by \citet{Navin2015}. \cite{Marino2014} found that the halo stars had \bafe\ abundances similar to the tidal member stars of the bright SGB, with an average $\bafe = 0.5$ (though with large error bars). They proposed that these stars therefore belong to the first generation of stars to form in the cluster, as this generation of stars would be less centrally concentrated than the second generation stars. We note that their sample has a specific selection function: they were aiming to maximize their sample of extratidal stars and so targeted the densely populated SGB.

In contrast to the \bafe\ abundance, there is no obvious correlation between our populations and the abundance of silicon or magnesium, except for the extreme population. For magnesium, this population is generally less abundant than the other populations. Also of interest with magnesium is that the anomalous population has almost the same \mgfe\ for all of its stars. No other population nor chemical element has such homogeneity in our sample of stars.

Overall, the picture that emerges of our populations is that the first and second generations of stars are relatively well separated in most elements (except for magnesium and silicon). However, although distinct in their CN and CH index strengths, the second generation stars (the intermediate, extreme and anomalous populations) are not separated in chemical abundances.

The ability to divide stars into different populations is a strength of low resolution work, but the lack of detailed information is a weakness. This is why both low resolution and high resolution approaches are useful.

\subsection{\cno}\label{sec:cno}
\begin{figure}
	\includegraphics[width=\columnwidth]{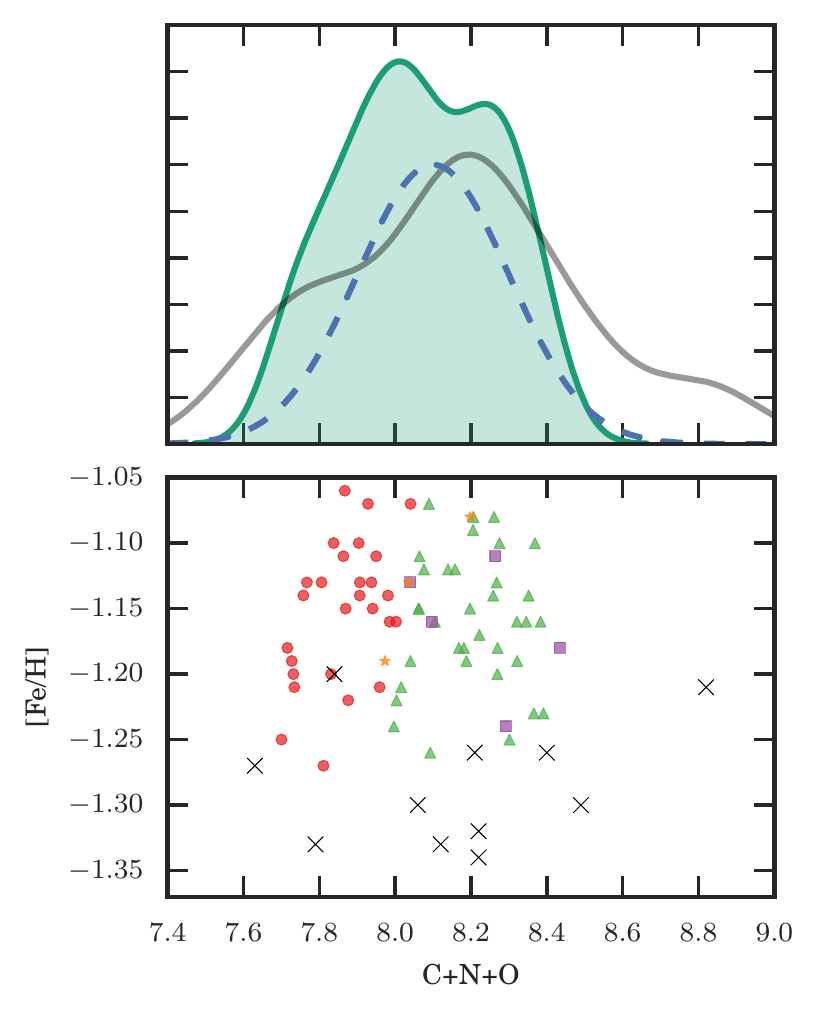}
    \caption{Top: KDE distribution of the \cno\ abundances determined for NGC~1851. The green shaded distribution is for this work, and the black line is for \citet{Yong2009}. The dashed blue line is a Gaussian with $\sigma=0.2$, the median uncertainty of the \cno\ values for this work. Bottom: The raw distribution of the \cno\ abundances. Data marked with crosses are from \citet{Yong2009} and the other symbols are as defined in Figure \ref{fig:j_dCN_kde_all}. The offset in \feh\ between \citet{Yong2009} and \citet{Carretta2011} is $0.14\pm0.03$ dex \citep[and was noted by][]{Carretta2011}.}
    \label{fig:vmag_cno}
\end{figure}

One of the key results from \cite{Yong2014b} was that NGC~1851 had a larger spread in the \cno\ abundance of its stars than would be expected from measurement uncertainties. In Figure \ref{fig:vmag_cno} we plot our \cno\ abundances along with those of \cite{Yong2014b}. The \cno\ sum is dominated by the nitrogen abundance, so it unsurprising, given the strong grouping of elemental abundances with the CN populations, that we find that the stars with low or high \cno\ also have low or high \dcnblue. It is interesting that the anomalous stars (strong in both CN and CH) do not stand out in their \cno. This is a consequence of their average (for the cluster stars) \ofe.

In Figure \ref{fig:vmag_cno} the \cno\ abundances have a larger spread than their median uncertainties. This is line with the result of \citet{Yong2014b} than there is in fact a spread in \cno\ for NGC~1851. With the exclusion of the star with the extreme \cno\ from \citet{Yong2014b}, it would appear that our study and theirs have identified very similar ranges of \cno. This is encouraging as they used much higher resolution spectra ($R\sim32000$) and used NH rather than CN features. This meant that their nitrogen abundances are not affected by molecular equilibria with carbon and oxygen.

Our \cno\ has a $\sigma = 0.2$, compared to \cite{Yong2014b} who found $\sigma = 0.34\pm0.08$. This spread is driven in part by what appears to be an outlier star when compared to our sample, with a $\cno=8.82$. We do not find any stars with $\cno>8.5$. If we exclude that star, then \cite{Yong2014b} has a $\sigma = 0.26$, which would bring their results closer to our spread in \cno.

\subsection{Radial distribution}\label{sec:radial}

\begin{figure}
	\includegraphics[width=\columnwidth]{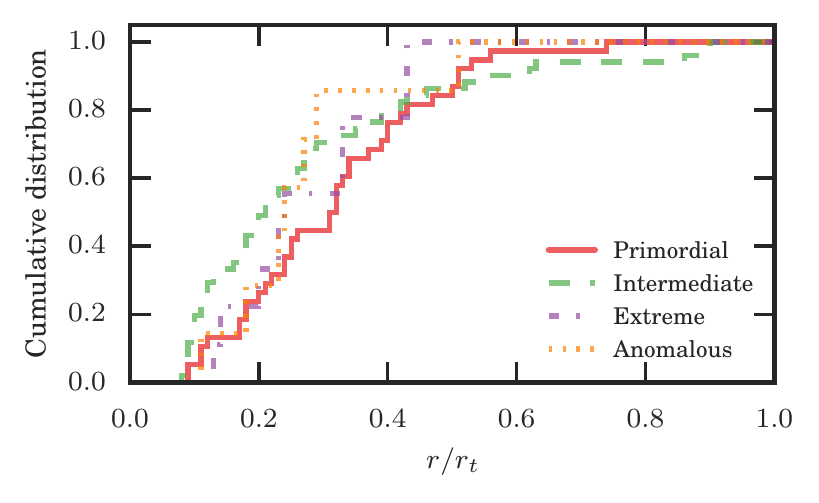}
    \caption{The radial distribution of the CN-CH populations as defined in Figure \ref{fig:j_dCN_kde_all}. The x axis is the distance from the cluster centre in units of the tidal radius ($r_t$). The second generation stars are somewhat more centrally concentrated that the primordial population. For radial distances larger than $0.6r_t$ there is only one primordial star, which is why its cumulative distribution quickly reaches unity.}
    \label{fig:dCN_radial}
\end{figure}

The radial distribution of the various populations of stars in globular clusters provides evidence about the concentration of gas from which the stars of each population formed. Theoretical models for globular cluster formation predict that the low-velocity gas ejected from AGB stars should accumulate in the centre of the cluster gravitational well. It would be from this cool, centrally-concentrated gas that the second generation of stars would form.

Observational results generally agree with this: the second generation stars are more centrally concentrated in a number of clusters \citep[NGC~2808;][]{Carretta2015b} \citep[NGC~362;][]{Carretta2013} \citep[47~Tuc][]{Cordero2014}. However, there is at least one exception to this: M15, which \cite{Larsen2015} found to have a more centrally concentrated primordial population. Finally, \cite{Dalessandro2014} found that NGC~6362 had no difference between the radial distributions of its stellar populations, suggesting that its populations have had enough time to radially homogenize. \citet{Cummings2014} used photometry of NGC~1851 and found that it was only the main sequence stars and not the RGB stars that showed any difference in the radial distribution of their photometrically-derived populations, with the redder MS population having a higher central concentration.

Using the CN-CH populations, we find that the second generation stars appear to be more centrally concentrated than the primordial population (Figure \ref{fig:dCN_radial}). This would be consistent with previous studies of globular clusters,  which have found that the second generation of stars are more centrally concentrated that the first generation, but not with the specific result from \citet{Cummings2014}.

For the radial distance range where there are a reasonable number of stars in each population $(0.07<r/r_t<0.6)$, a two-sample Kolmogorov-Smirnov test yields a p-value of 0.06 and a KS statistic of 0.25. Such a result is on the cusp of allowing us to reject the null hypothesis that the radial distributions are drawn from the same sample.

\subsection{CN-richest populations}\label{sec:anomalous}

\begin{figure}
	\includegraphics[width=\columnwidth]{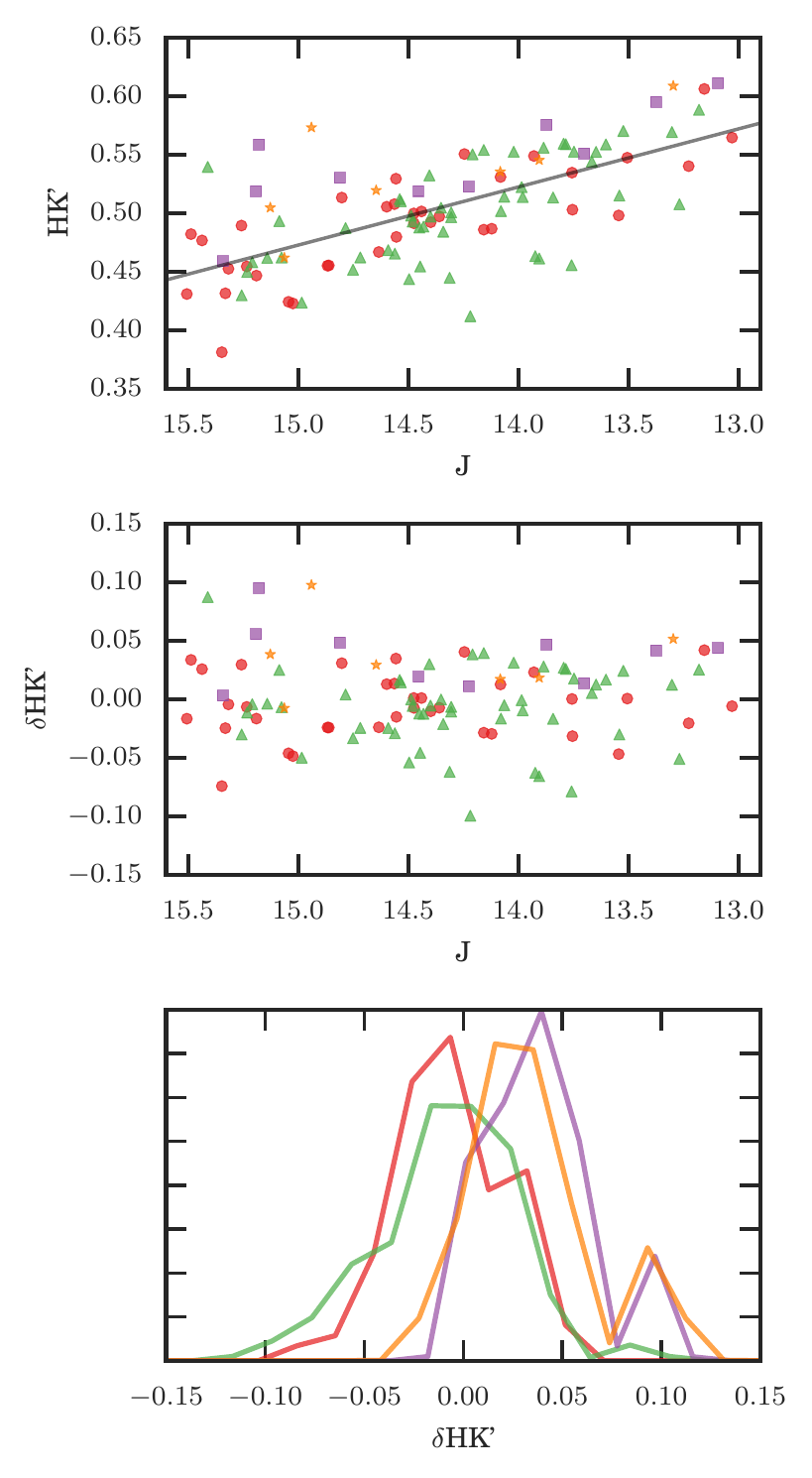}
    \caption{Top: The \hkdash\ indices of stars coded with their CN-CH populations. The symbols are as defined in Figure \ref{fig:j_dCN_kde_all}. The straight line is the best fit to the data. Middle: The \hkdash\ values adjusted using the straight line from the top panel (to give $\delta\hkdash$). Bottom: KDE of the $\delta\hkdash$ of the four populations. The key finding is that the two most CN-rich populations both generally have stronger Calcium H \& K lines. These lines are a metallicity indicator.}
    \label{fig:j_hk_pops}
\end{figure}

Unlike many other clusters, NGC~1851 has been previously found not to exhibit a correlation between its CH and CN spectral indices. Except for the anomalous population, the populations show about the same range of \sch, though with a slight preference for the primordial population to be more CH-rich than the intermediate population.

The normal interpretation of the CN-CH anti-correlation is that it is driven by an anti-correlation between [C/Fe] and [N/Fe]. There is not a clear anti-correlation between CN and CH band strengths in NGC 1851, as can be seen in Figure \ref{fig:j_dCN_kde_all}. However, it is clear from Figures \ref{fig:teff_cfe} and \ref{fig:chem_populations} that the stars in our primordial population are both more carbon-rich and nitrogen-poor than the intermediate population. This illustrates the limitations of using molecular band strengths to infer abundances, as also shown by \citet{Boberg2016}.

We can use \hkdash\ to investigate the metallicity of our stars. In Figure \ref{fig:j_hk} it is clear that there is a correlation between the luminosity and the \hkdash\ of the cluster stars. In a manner similar to that used to correct the \cnblue\ for this effect, a straight line was fitted and the \hkdash\ adjusted (Figure \ref{fig:j_hk_pops}). The two most CN-rich populations are enhanced in their \hkdash\ index for their luminosity.

\section{Concluding remarks}
This paper presents new determinations for \cfe, \nfe, and related spectral indices for a sample of red giant branch stars in the globular cluster NGC~1851 with complementary abundance data from the literature. Using the CN and CH spectral indices we find that the stars in this cluster divide into four groups, similar to the finding of \citet{Campbell2012}. The majority of the stars belong to two main groups that exhibit the typical globular cluster anticorrelations (C-N, O-Na, Mg-Al) and also show variations in Si and Ba abundance. A third more CN-rich group sits further along the abundance anticorrelations than the other two, and the group with the strongest CN absorption also shows unusually high abundances of C and Ba, suggesting that its nucleosynthetic influences are different, or at least broader, than the other groups.

Combined analyses of bandstrengths and abundance measurements are uncommon, generally because they require two independent data sets. Fortunately, there are recent examples \citep{Smith2015a,Smith2015} for 47~Tuc and M71 to consider alongside our results. These studies find that in both clusters: (i) there is an anticorrelation between a CN bandstrength indicator and \ofe; (ii) there is a positive correlation between CN and \nafe; and (iii) these relations are quite similar despite the significant differences in mass and central concentration of the two clusters. However, the range of \ofe\ is larger in 47~Tuc than in M71, something that has been shown to correlate with higher cluster mass \citep{Carretta2010b}. In addition, the fraction of CN-strong stars is higher, and those CN-strong stars are relatively more centrally concentrated in 47~Tuc. In both clusters, the CN data can be used to divide the stars into fairly clear-cut groups, but those groups are not as well-separated in \ofe\ or \nafe. This is quite similar to our results in NGC~1851, and can be seen clearly in Figure \ref{fig:chem_populations}.

We confirm that indicators based on both high- and low-resolution spectroscopy can be used to identify chemical complexity in globular clusters, though each has its strengths and weaknesses. CN and CH indices make the presence of complex populations quite clear, but they do not carry information about abundance variations in other elements (e.g., the extent of the Mg-Al anticorrelation or whether there are Si variations). There are indications that the O-N and Mg-Al anticorrelations can be divided into discrete groups similar to the CN groups \citep{Carretta2015b}, and that they are not intrinsically smoothly populated. Since the nucleosynthetic processes that produce the C-N, O-Na and Mg-Al reactions operate at different temperatures, it is critical to consider all of these elements when searching for the combination of stellar feedback sources responsible for globular cluster abundance anomalies.

We are also able to confirm that the stars proposed as extratidal cluster stars by \citet{Navin2015}, based on photometry and radial velocity, also carry the light-element chemical tags of NGC 1851. This indicates that they did originate in the cluster and have subsequently been lost, or possibly that they belong to a larger cluster halo left over from a progenitor dwarf galaxy (as suggested by \citealt{Marino2014}). Assuming that they are extratidal members we have estimated a cluster destruction rate that is comparable to model estimates but it does rely on several assumptions that could cause it to become much higher or much lower.

In addition, we confirm a range in total \cno\ abundance that correlates with CN bandstrength and Na abundance, in agreement with \citet{Yong2009,Yong2014b}. This is a strong indication that lower-mass AGB stars must have contributed to the chemical enrichment of NGC 1851. This abundance behaviour, along with the metallicity range measured by \citet{Carretta2010}, sets NGC 1851 apart from ordinary globular clusters, and may shed light on its formation and enrichment history.

Finally, we consider the evolutionary changes in C abundance driven by deep mixing. We find that our data fit well to a model with two parallel trends, consistent with expectations for a population of stars with primordial variations in \cfe\ followed by evolutionary mixing as stars ascend the giant branch. The mixing rate we find, in dex per V magnitude, has a reasonable agreement with the empirical study of \citet{Martell2008a} and the theoretical work of \citet{Angelou2012}. Our confidence in this result is tempered by the limited number of pre-bump stars in our data set, which makes it difficult to place the level of carbon abundance before the onset of deep mixing. There could be a mild metallicity range in NGC~1851, with \feh\ ranging from $-1.25$ to $-1.05$ \footnote{Though this is contested for NGC~1851 \citep{Yong2014b}. It has been also found that some metallicity spreads are the result of differences in \logg\ derived spectroscopically and photometrically \citep{Mucciarelli2015}.}. The rate of deep mixing is sensitive to metallicity, but over this narrow metallicity range we would not expect a dramatic difference. We calculate that, for two stars with metallicities of $-1.25$ and $-1.05$ and identical initial \cfe, the difference between their \cfe\ by the time they reach the bright end of our sample ($1.4$ magnitudes brighter than the luminosity function bump) will only be $0.11$ dex, which is similar to our measurement error for [C/Fe]. A data set including stars up to the tip of the giant branch, deriving carbon abundances more precisely from higher-resolution spectra, ought to be able to detect different rates of deep mixing between the most metal-poor and metal-rich stars in NGC 1851.

Globular clusters are no longer viewed as simple stellar populations, but they remain excellent laboratories for stellar evolution. Although we have known for quite some time that deep mixing occurs in stars brighter than the luminosity function bump, it is still unclear what drives the mixing process \citep[e.g.,][]{Charbonnel2010,Eggleton2008} and what the best way is to represent it in stellar evolution models \citep[e.g.,][]{Angelou2012}. Large-scale specroscopic survey projects ought to be able to clarify this situation by providing information on key indicators of deep mixing such as Li, C, N, and $^{12}$C/$^{13}$C, determined homogeneously, for red giants all along the giant branch in globular clusters across a broad range in metallicity.

The spectroscopic study of globular clusters has been dominated by the analysis of light element abundance variations. The mechanism responsible for imprinting these variations at some level in nearly all known globular clusters is presently unclear. Regardless of their origin, light element abundances can act as an effective chemical tag denoting that a star formed in a globular cluster, whether it is just slightly outside the tidal radius (as in \citealt{Navin2015}) or it has been completely lost to the halo field \citep[as in][]{Martell2010,Martell2011}.

In conclusion, in this work we have explored the following aspects of this complicated globular cluster:
\begin{itemize}
	\item There are four populations of stars in NGC~1851 based upon their CN and CH spectral indices. Most of the stars belong to the populations with the primordial and intermediate populations. The anomalous population also shows CH enhancement that is not observed in the other populations.
	\item The light element abundance patterns are complex with regard to these populations. It is important to combine the results of both low and high resolution spectra.
	\item Barium was the only element, apart from carbon and nitrogen, that tracked with increasing CN.
	\item The extreme and anomalous populations have higher calcium \hkdash\ index values than the other two populations, and appear to be redder on the CMD.
	\item The primordial and intermediate populations of stars exhibit similar rates of carbon astration as the stars ascend the giant branch.
	\item The previous cluster membership of four extratidal stars identified by \citet{Navin2015} has been confirmed based upon their CN and CH spectral index strengths.
\end{itemize}

\section*{Acknowledgements}
We thank the referee, Elena Pancino, for her insightful comments that improved the manuscript.

SLM acknowledges financial support from the Australian Research Council through DECRA Fellowship DE140100598.

The data in this paper were based on observations obtained at the Australian Astronomical Observatory as part of AAO2012B/18. We thank Daniela Carollo who was one the initiators of this project.

The following software and programming languages made this research possible: \textsc{2dfdr} \citep[version 6.14;][]{AAOSoftwareTeam2015}, the 2dF Data Reduction software; Python (version 3.5); Astropy \citep[version 1.2.1;][]{Robitaille2013}, a community-developed core Python package for Astronomy; pandas \citep[version 0.18.1;][]{McKinney2010}; Tool for OPerations on Catalogues And Tables \citep[\textsc{topcat}, version 4.3-3;][]{Taylor2005}.

This publication makes use of data products from the Two Micron All Sky Survey, which is a joint project of the University of Massachusetts and the Infrared Processing and Analysis Center/California Institute of Technology, funded by the National Aeronautics and Space Administration and the National Science Foundation.

The national facility capability for SkyMapper has been funded through ARC LIEF grant LE130100104 from the Australian Research Council, awarded to the University of Sydney, the Australian National University, Swinburne University of Technology, the University of Queensland, the University of Western Australia, the University of Melbourne, Curtin University of Technology, Monash University and the Australian Astronomical Observatory. SkyMapper is owned and operated by The Australian National University's Research School of Astronomy and Astrophysics. The survey data were processed and provided by the SkyMapper Team at ANU. The SkyMapper node of the All-Sky Virtual Observatory is hosted at the National Computational Infrastructure (NCI).




\bibliographystyle{mnras}



\bsp	
\label{lastpage}
\end{document}